# Generalized Bagley-Torvik Equation and Fractional Oscillators

Mark Naber
*Monroe County Community College*

Lucas Lymburner
*Monroe County Community College*

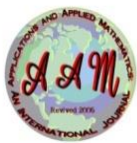

# Generalized Bagley-Torvik Equation and Fractional Oscillators

## Mark Naber and Lucas Lymburner

Department of Science and Mathematics
Monroe County Community College
1555 S. Raisinville Rd.
Monroe, Michigan, 48161-9746
mnaber@monroeccc.edu;

## Abstract

In this paper the Bagley-Torvik Equation is considered with the order of the damping term allowed to range between one and two. The solution is found to be representable as a convolution of trigonometric and exponential functions with the driving force. The properties of the effective decay rate and the oscillation frequency with respect to the order of the fractional damping are also studied. It is found that the effective decay rate and oscillation frequency have a complex dependency on the order of the derivative of the damping term and exhibit properties one might expect of a thermodynamic Equation of state: critical point, phase change, and lambda transition.

**MSC (2010) No:** 26A33, 34A08, 34A25, 34K11, 70J99

## 1. Introduction

In this paper a generalized Bagley-Torvik Equation is considered. The Bagley-Torvik Equation was originally derived to model a thin plate immersed in a Newtonian fluid and connected to a spring (Bagley and Torvik (1984), Podlubny (1999)). If a thin plate of mass $m$ and surface area $A$ is immersed in a Newtonian fluid with viscosity $\mu$ and density $\rho$ and attached to a spring with spring constant $K$ the following Equation describes the motion of the plate due to a force $f(t)$

164

$$mD_t^2 y(t) + 2\sqrt{\mu\rho}A_0 D_t^{3/2} y(t) + Ky(t) = f(t), \quad (1)$$
$$y(0) = y'(0) = 0.$$

For this problem the fluid and the plate are initially at rest and the plate is being driven by a known force $f(t)$. The shear stress of the fluid is the physical reason the fractional derivative appears in the Bagley-Torvik Equation (Bagley and Torvik (1984)). This Equation has been studied numerically (Podlubny (1999), Diethelm and Ford (2002)) and analytically (Ray and Bera (2005), Labecca et al. (2015), Bansal and Jain (2016)). The known analytical solutions are expressed in terms of an infinite series of Mittag-Leffler functions.

To consider a more general problem, so as to compare with the results of an earlier paper Naber (2010), and to offer an insight into the damped harmonic oscillator, the following Equation will be considered (generalized Bagley-Torvik Equation, GBT)

$$D_t^2 y(t) + \lambda_0 D_t^\nu y(t) + \omega^2 y(t) = f(t), \quad (2)$$
$$y(0) = y_0, \quad y'(0) = y_1,$$

with $1 < \nu < 2$ and $\lambda$ and $\omega$ real and positive (as they would be for the specific case of the Bagley-Torvik Equation). A similar Equation arises in the study of certain seismic models; see Equation (3) of Koh and Kelly (1990). Equation (2) will not be solved using a series of Mittag-Leffler functions, but rather, it will be solved directly with the Laplace inversion integral. This will allow the oscillating component of the solution to be expressed in terms of sine and cosine functions (and their convolutions) and allow for some interesting observations concerning the effective decay rate and oscillation frequency with respect to the parameters ($\lambda$, $\omega$, and $\nu$) of the GBT Equation. Note that the units of the damping constant, $\lambda$, will change with the order of the derivative $\nu$. Due to the non-local nature of the fractional derivative it is not possible to rescale the time coordinate and make the Equation dimensionless. For a non-Newtonian fluid, the order of the derivative on the damping term would not be expected to be $\nu = 3/2$ as the shear stress is much more complicated. See for example the off-diagonal terms of Equation (1) and the discussion that follows in Rajagopal and Na (1983).

The Caputo formulation of the fractional derivative will be used in this study (Podlubny (1999)). Recall that the Laplace transform of the Caputo fractional derivative is (for $1 < \nu < 2$),

$$\mathcal{L}(^C_0 D_t^\nu f(t)) = s^\nu F(s) - s^{\nu-1} f(0) - s^{\nu-2} f'(0). \quad (3)$$

The Caputo fractional derivative is found to be more useful for physical problems than other formulations of fractional derivatives due to the appearance of the physical initial conditions, $f(0)$ initial position, and, $f'(0)$ initial velocity, in the Laplace transform.

In section 2 the solution to the GBT Equation is found by means of Laplace transform. The solution presented in this paper is somewhat easier to interpret than those given by an infinite sum of Mittag-Leffler functions in that it is expressed in terms of sine, cosine, and exponential functions, and their convolutions. In section 3 the dependence of the effective decay rate and oscillation frequency on $\nu, \lambda,$ and $\omega$ is explored. An analogy with a thermodynamic Equation of state is made. The Equation relating the two dependent variables to the three independent variables has many properties one might expect from an Equation of state, including phase transitions and lambda transitions. It is found that there are 15 types of solution curves to this Equation of state. Representative graphs for each possibility are presented and discussed.

## 2. Constructing a Solution

Consider the Laplace transform of Equation (2),

$$s^2 Y(s) - s y_0 - y_1 + \lambda(s^\nu Y(s) - s^{\nu-1} y_0 - s^{\nu-2} y_1) + \omega^2 Y(s) = F(s), \qquad (4)$$

or,

$$Y(s) = \frac{F(s)}{s^2 + \lambda s^\nu + \omega^2} + \frac{(s y_0 + y_1)(1 + \lambda s^{\nu-2})}{s^2 + \lambda s^\nu + \omega^2}. \qquad (5)$$

If Equation (5) is inverted using a contour integral a branch cut is needed on the negative real axis due to the fractional exponents on the complex variable $s$. Hence, a Hankel contour will be used. This contour starts at $\gamma - i\infty$ goes vertically up to $\gamma + i\infty$ (where $\gamma$ is chosen so that all poles will lie to the left of the vertical contour line) then, travels in a quarter circle arc (to the left) to just above the negative real axis (i.e., $-\infty$). Then, the contour has a cut that goes into the origin (following the negative real axis), around the origin in a clockwise sense (to just below the negative real axis) then, back out to $-\infty$. The contour is completed with another quarter circle arc to $\gamma - i\infty$.

To evaluate this integral the poles need to be found. The poles are the roots of

$$s^2 + \lambda s^\nu + \omega^2 = 0. \qquad (6)$$

Equation (6) will be solved similarly to Naber (2010). The fractional exponent on the second term effectively makes Equation (6) transcendental with closed form solutions for only a few special cases of $\nu$. Let $s = r e^{i\theta}$ then, Equation (6) breaks into two Equations, real and imaginary,

$$\begin{aligned} r^2 \cos(2\theta) + \lambda r^\nu \cos(\nu\theta) + \omega^2 &= 0, \\ r^2 \sin(2\theta) + \lambda r^\nu \sin(\nu\theta) &= 0. \end{aligned} \qquad (7)$$

Note that $1 < \nu < 2$. Could there be a solution on the positive real axis? No, in this case $\theta = 0$ and the first Equation of Equation (7) would be the sum of three positive non-zero terms, which would never be zero. Could there be a solution on the negative real axis? No, in this case $\theta = \pi$ and the second Equation of Equation (7) would never be satisfied. Using similar arguments we can show that there are no solutions on the positive or negative imaginary axes. It can also be shown that no solutions are in the right half plane (both terms of the second Equation would always be positive). If there are solutions they should be in pairs, complex conjugates, with $\frac{\pi}{2} < \theta < \pi$ and $-\pi < \theta < -\frac{\pi}{2}$. To attempt to find a solution first solve the second Equation of Equation (7) for $r$ and substitute this into the first Equation (only look for positive $\theta$ values),

$$\left(-\lambda \frac{\sin(v\theta)}{\sin(2\theta)}\right)^{2/(2-v)} \cos(2\theta) + \lambda \left(-\lambda \frac{\sin(v\theta)}{\sin(2\theta)}\right)^{v/(2-v)} \cos(v\theta) + \omega^2 = 0. \tag{8}$$

Equation (8) can be simplified to,

$$\left(\frac{(\sin(v\theta))^v}{(\sin(2\theta))^2}\right)^{1/(2-v)} \sin((2-v)\theta) = \left(\frac{\omega}{\lambda^{1/(2-v)}}\right)^2. \tag{9}$$

Now it needs to be seen if there is a $\theta$ value that will satisfy Equation (9). For this Equation to be true $\sin(v\theta)$ and $\sin((2-v)\theta)$ need to be positive. This will only happen on the restricted domain $\frac{\pi}{2} < \theta < \frac{\pi}{v}$. Now the question becomes, on this restricted domain can we pick a $\theta$ value that will make the left-hand side of Equation (9) as large or as small as we wish? Thus, ensuring that no matter what the values of $\omega$, $\lambda$, and $v$ we are given we can always find a $\theta$ value that will satisfy Equation (9). Consider the two limits

$$\lim_{\theta \to \frac{\pi}{2}^+} \left(\frac{(\sin(v\theta))^v}{(\sin(2\theta))^2}\right)^{1/(2-v)} \sin((2-v)\theta) = \infty, \tag{10}$$

$$\lim_{\theta \to \frac{\pi}{v}^-} \left(\frac{(\sin(v\theta))^v}{(\sin(2\theta))^2}\right)^{1/(2-v)} \sin((2-v)\theta) = 0. \tag{11}$$

Since the left hand side of Equation (9) is continuous in $\theta$ and we have the two limits above, Equations (10) and (11), it is guaranteed that there will be at least one solution to Equation (9) and there will be at least two poles for the residue calculation. If we can show that the left-hand side of Equation (9) decreases monotonically in $\theta$ over the restricted domain then, we know that there will be only one solution to Equation (9), and thus, only two poles in the residue calculation. To show that the left-hand side of Equation (9) decreases monotonically in $\theta$ we need to show that the derivative of the left-hand side of Equation (9) with respect to $\theta$ is always negative, i.e.,

$$\frac{\partial}{\partial \theta} \left\{ \left(\frac{(\sin(v\theta))^v}{(\sin(2\theta))^2}\right)^{1/(2-v)} \sin((2-v)\theta) \right\} < 0. \tag{12}$$

Computing the derivative, doing some algebra, and dividing out over all factors that have a constant sign gives,

$$v^2(\sin(2\theta))^2 - 4v\sin(2\theta)\sin(v\theta)\cos((2-v)\theta) + 4(\sin(v\theta))^2 > 0. \tag{13}$$

On the restricted domain $0 < \cos((2-v)\theta) < 1$. This reduces Equation (13) to

$$\nu^2(\sin(2\theta))^2 - 4\nu\sin(2\theta)\sin(\nu\theta) + 4(\sin(\nu\theta))^2 > 0, \tag{14}$$

which can now be factored into a perfect square and prove the assertion made in Equation (12),

$$(\nu\sin(2\theta) - 2\sin(\nu\theta))^2 > 0. \tag{15}$$

Hence, the left-hand side of Equation (9) will decrease monotonically on the restricted domain with the upper bound being $\infty$ and the lower bound being 0. To summarize, it has just been shown that there is always one solution, with a positive angle, to Equation (9) and this solution must be such that $\frac{\pi}{2} < \theta < \frac{\pi}{\nu}$. Consequently, there will be two poles for the residue calculation, and they will be complex conjugates of each other. Notice that for the fractionally damped Equation repeated roots are not possible. Repeated roots can only happen when the order of the derivative becomes one ($\nu = 1$).

Now that the question of the poles has been settled the solution to Equation (2) can be generated. Denote the two poles as

$$s_{1,2} = \beta \pm i\sigma = re^{\pm i\theta}, \tag{16}$$

where $\beta$ and $\sigma$ are determined from the $r$ and $\theta$ values that satisfy Equation (9) in the usual way, $r = \sqrt{\beta^2 + \sigma^2}$ and $\tan(\theta) = \sigma/\beta$. Note that $\beta$ is negative, the two solutions are in the second and third quadrants and $s_1$ is the complex conjugate of $s_2$. The poles are of order one and the residue is given by,

$$\lim_{s \to s_1} \frac{(s - s_1)(sy_0 + y_1)(1 + \lambda s^{\nu-2})e^{st}}{s^2 + \lambda s^\nu + \omega^2} + \lim_{s \to s_2} \frac{(s - s_2)(sy_0 + y_1)(1 + \lambda s^{\nu-2})e^{st}}{s^2 + \lambda s^\nu + \omega^2}, \tag{17}$$

$$= \frac{(s_1 y_0 + y_1)(1 + \lambda s_1^{\nu-2})e^{s_1 t}}{2s_1 + \lambda s_1^{\nu-1}} + \frac{(s_2 y_0 + y_1)(1 + \lambda s_2^{\nu-2})e^{s_2 t}}{2s_2 + \lambda s_2^{\nu-1}}. \tag{18}$$

After some algebra this can be reduced to,

$$\frac{(ay_0 + by_1)\cos(\sigma t) + (cy_0 + dy_1)\sin(\sigma t)}{4r^2 + 4\nu\lambda r^\nu \cos((2-\nu)\theta) + (\nu\lambda r^{\nu-1})^2} e^{\beta t}, \tag{19}$$

where,

$$a = 2\left(2r^2 + \nu\lambda^2 r^{\nu-1} + \lambda(2 + \nu)r^\nu \cos((2-\nu)\theta)\right), \tag{20}$$

$$b = 2\left(\beta(\nu\lambda^2 r^{\nu-2} + 2) + 2\lambda r^{\nu-1}\cos((\nu-3)\theta) + \nu\lambda r^{\nu-1}\cos((\nu-1)\theta)\right), \tag{21}$$

$$c = 2\lambda(\nu - 2)r^\nu \sin((\nu-2)\theta), \tag{22}$$

$$d = 2\Big(\sigma(\nu\lambda^2 r^{\nu-2} + 1) - 2\lambda r^{\nu-1}\sin((\nu-3)\theta) + \nu\lambda r^{\nu-1}\sin((\nu-1)\theta)\Big). \quad (23)$$

For the contour integral the only contributions come from the paths along the negative real axis. This is given by the following integral

$$\frac{\lambda}{\pi}\int_0^\infty \frac{(y_1 - Ry_0)\Big((R^\nu + \omega^2 R^{\nu-2})\sin((\nu-2)\pi) - R^\nu \sin(\nu\pi)\Big)e^{-Rt}}{(R^2+\omega^2)^2 + 2\lambda R^\nu(R^2+\omega^2)\cos(\nu\pi) + (\lambda R^\nu)^2} dR. \quad (24)$$

Define a function $C(y_0, y_1, \nu, \lambda, \omega, R)$, such that Equation (24) can be written as,

$$\int_0^\infty C e^{-Rt} dR. \quad (25)$$

The homogeneous solution to Equation (2) is then, Equation (24), or (25), subtracted from Equation (19). This may look overly complicated but the solution does have the general form of

$$y(t) = e^{\beta t}\big(A\cos(\sigma t) + B\sin(\sigma t)\big) - \int_0^\infty C e^{-Rt} dR, \quad (26)$$

where,

$$A = \frac{(ay_0 + by_1)}{4r^2 + 4\nu\lambda r^\nu \cos((2-\nu)\theta) + (\nu\lambda r^{\nu-1})^2}, \quad (27)$$

$$B = \frac{(cy_0 + dy_1)}{4r^2 + 4\nu\lambda r^\nu \cos((2-\nu)\theta) + (\nu\lambda r^{\nu-1})^2}. \quad (28)$$

Notice that Equation (24) goes to zero if $\nu$ goes to one or two, i.e., Equation (2) goes to its non-fractional limits the decay function (Equation (25)) goes away, as expected. The damping factor $e^{\beta t}$ is similar to the damping factor for the non-fractional case, $e^{-\lambda t/2}$. Notice that since the poles, for the residue calculation, have non-zero imaginary and non-zero real parts we will not have the same three distinct cases as we did for the non-fractional case (critically-damped, over-damped, and under-damped).

In Naber (2010) a similar solution was found for the homogeneous version of Equation (2) of this paper. The only difference was in the location of the poles for the inverse Laplace transform integral. For Equation (2) with $0 < \nu < 1$ it was found that $\frac{\pi}{2} < \theta < \frac{\pi}{2-\nu}$. For Equation (2) with $1 < \nu < 2$ it was found that $\frac{\pi}{2} < \theta < \frac{\pi}{\nu}$.

The solution to the full inhomogeneous problem can now be written down in terms of a convolution integral. Consider the following expression

$$2e^{\beta t}\cos(\sigma t)\frac{2r\cos(\theta) + \nu\lambda r^{\nu-1}\cos((\nu-1)\theta)}{4r^2 + 4\nu\lambda r^\nu\cos((2-\nu)\theta) + (\nu\lambda r^{\nu-1})^2}$$
$$+2e^{\beta t}\sin(\sigma t)\frac{2r\sin(\theta) + \nu\lambda r^{\nu-1}\sin((\nu-1)\theta)}{4r^2 + 4\nu\lambda r^\nu\cos((2-\nu)\theta) + (\nu\lambda r^{\nu-1})^2} \qquad (29)$$
$$-\frac{\lambda}{\pi}\int_0^\infty \frac{\sin(\nu\pi)\, R^\nu e^{-Rt}}{(R^2+\omega^2)^2 + 2\lambda R^\nu(R^2+\omega^2)\cos(\nu\pi) + (\lambda R^\nu)^2}\, dR\,.$$

Define $\tilde{A}$, $\tilde{B}$, and $\tilde{C}$ so that Equation (29) can be written as

$$\tilde{A}e^{\beta t}\cos(\sigma t) + \tilde{B}e^{\beta t}\sin(\sigma t) - \int_0^\infty \tilde{C}e^{-Rt}\,dR\,. \qquad (30)$$

The general solution is given by

$$y(t) = Ae^{\beta t}\cos(\sigma t) + Be^{\beta t}\sin(\sigma t) - \int_0^\infty Ce^{-Rt}\,dR$$
$$+ \int_0^t f(t-u)\left(\tilde{A}e^{\beta u}\cos(\sigma u) + \tilde{B}e^{\beta u}\sin(\sigma u) - \int_0^\infty \tilde{C}e^{-Ru}\,dR\right)du. \qquad (31)$$

This solution is expressed completely in terms of the exponential, sine, and cosine functions and their convolutions with the forcing term. Given this form of solution $\beta$ and $\sigma$ can be thought of as effectively defining a decay rate and oscillation frequency, especially if the driving force is a delta function.

Now consider specifically Equation (1). In this case $\nu = 3/2$, $\lambda = \frac{2A\sqrt{\mu\rho}}{m}$, and $\omega^2 = \frac{K}{m}$. The solution to the Bagley-Torvik Equation is given by,

$$y(t) = \int_0^t f(t-u)\left(\tilde{A}e^{\beta u}\cos(\sigma u) + \tilde{B}e^{\beta u}\sin(\sigma u) - \int_0^\infty \tilde{C}e^{-Ru}\,dR\right)du. \qquad (32)$$

## 3. Oscillation Frequency and Decay Rate

The frequency of the oscillation component of the solution is, $\sigma = \Im(s)$ (the imaginary part of the pole in the upper half plane, the subscript will be suppressed) and the decay rate is $\beta = \Re(s)$ (the real part of the pole in the upper half plane). The Equation defining these two variables can be thought of as an Equation of state with two dependent variables $\sigma$ and $\beta$ and three independent variables $\nu$, $\lambda$, and $\omega$ (e.g. pages 40-42 of Sears and Salinger (1975))

$$s = \beta + i\sigma, \qquad (33)$$

$$s^2 + \lambda s^\nu + \omega^2 = 0, \qquad (34)$$

$$\sigma = \sigma(\nu, \lambda, \omega),$$

$$\beta = \beta(\nu, \lambda, \omega), \tag{35}$$

$$0 \leq \nu \leq 2, \quad \text{and,} \quad 0 \leq \lambda, \omega < \infty.$$

As noted earlier the Equation of state (Equation (34) or Equation (6)) is effectively transcendental so it cannot be solved explicitly for the oscillation frequency or the decay rate except for a few trivial cases of the independent variables;

$$\nu = 0, \quad \Rightarrow \quad \beta = 0, \quad \sigma = \sqrt{\lambda + \omega^2}, \tag{36}$$

$$\lambda = 0, \quad \Rightarrow \quad \beta = 0, \quad \sigma = \omega, \tag{37}$$

$$\nu = 2, \quad \Rightarrow \quad \beta = 0, \quad \sigma = \frac{\omega}{\sqrt{1+\lambda}}. \tag{38}$$

When $\nu = 1$ the usual three cases for a damped harmonic oscillator occur; over-damped, critically-damped, and under-damped. The frequency may be zero or non-zero, and the decay rate will be double valued for one case.

$$\begin{aligned}
\lambda > 2\omega, &\quad \Rightarrow \quad \beta = \frac{-\lambda \pm \sqrt{\lambda^2 - 4\omega^2}}{2}, \quad \sigma = 0, \\
\lambda = 2\omega, &\quad \Rightarrow \quad \beta = -\frac{\lambda}{2}, \quad \sigma = 0, \\
\lambda < 2\omega, &\quad \Rightarrow \quad \beta = -\frac{\lambda}{2}, \quad \sigma = \frac{\sqrt{4\omega^2 - \lambda^2}}{2}.
\end{aligned} \tag{39}$$

Three other explicit solutions can be found for $\nu = 1/2$, $\nu = 3/2$, and $\nu = 2/3$ using the formulas for quartic and cubic polynomial Equations. For $0 < \nu < 1$ or $1 < \nu < 2$ there will always be a non-zero frequency due to the location of the poles in the residue computation of the inverse Laplace transformation. Note that $\frac{\omega}{\sqrt{1+\lambda}} < \sqrt{\lambda + \omega^2}$ and $0 < \sqrt{\omega^2 - \lambda^2/4} \leq \sqrt{\lambda + \omega^2}$ (assuming $\lambda < 2\omega$).

In the non-fractional case ($\nu = 1$) increasing $\lambda$ causes the frequency of oscillation to become smaller, monotonically, until the critical cases are reached and the oscillation period becomes infinite (these are the critical and over-damped cases). In the fractional case the frequency of oscillation, $\sigma$, now depends on the order of the derivative, $\nu$, as well as $\lambda$ and $\omega$. One question that might asked is: how do the frequency and the decay rate change as the independent variables are changed? Some easily obtainable properties are listed below without proof.

$$\left.\frac{\partial \beta}{\partial \lambda}\right|_{\nu=0} = 0, \quad \left.\frac{\partial \beta}{\partial \lambda}\right|_{\nu=1} < 0, \quad \left.\frac{\partial \beta}{\partial \lambda}\right|_{\nu=2} = 0, \tag{40}$$

$$\frac{\partial \beta}{\partial \lambda} = 0, \quad \text{when,} \quad 2(1-\nu)\beta^2 + 2(1+\nu)\sigma^2 + (2-\nu)(\beta^2 + \sigma^2)^2 = \nu\omega^2, \tag{41}$$

$$\left.\frac{\partial \sigma}{\partial \lambda}\right|_{\nu=0} > 0, \qquad \left.\frac{\partial \sigma}{\partial \lambda}\right|_{\nu=2} < 0, \qquad \left.\frac{\partial \sigma}{\partial \lambda}\right|_{\sigma=0} = 0, \tag{42}$$

$$\left.\frac{\partial \beta}{\partial \omega}\right|_{\omega=0} = 0, \qquad \left.\frac{\partial \beta}{\partial \omega}\right|_{\nu=0} = 0, \qquad \left.\frac{\partial \beta}{\partial \omega}\right|_{\nu=2} = 0, \tag{43}$$

$$\frac{\partial \beta}{\partial \omega} = 0, \quad \text{when,} \quad \beta^2 + \sigma^2 = \frac{\nu\omega^2}{2-\nu}, \tag{44}$$

$$\frac{\partial \sigma}{\partial \omega} \geq 0 \quad \forall \; \nu, \lambda, \omega. \tag{45}$$

Some interesting observations arise when $\lambda$ and $\omega$ are fixed and the order of the fractional derivative is varied. The "Equation of state" (Equation (34)) now defines a curve in the complex plane. All curves start ($\nu = 0$) at $\left(0, i\sqrt{\lambda + \omega^2}\right)$, end ($\nu = 2$) at $\left(0, i\frac{\omega}{\sqrt{1+\lambda}}\right)$, and stay in the second quadrant. Recall that the fractional time derivatives are frequently interpreted as incorporating memory into the system (e.g. pages 88-90 of Podlubny (1999) see also Equation (2) of Koh and Kelly (1990)). Additionally, the decay rate can be double valued at $\nu = 1$, so, non-monotonic behavior with respect to the independent variable $\nu$ is to be expected.

It can be shown that there are 15 types of behavior for the decay rate and the frequency with respect to the order of the fractional derivative. As before, attention will be restricted to the upper half plane for $s$. These 15 designations are with respect to the initial and final slope of the frequency with respect to $\nu$ and the types of solutions to Equation (34) at the point $\nu = 1$.

The slope of the frequency, $\sigma$, and the decay rate, $\beta$, with respect to the order of the fractional derivative are given by the real and imaginary parts of (recall that $\omega$ and $\lambda$ are held fixed)

$$\frac{ds}{d\nu} = \frac{s(s^2 + \omega^2)\ln(s)}{(2-\nu)s^2 - \nu\omega^2}, \tag{46}$$

$$\frac{d\sigma}{d\nu} = \Im\left(\frac{ds}{d\nu}\right), \tag{47}$$

$$\frac{d\beta}{d\nu} = \Re\left(\frac{ds}{d\nu}\right). \tag{48}$$

Consider Equation (46) when $\nu = 0$,

$$\left.\frac{ds}{d\nu}\right|_{\nu=0} = \frac{(s^2 + \omega^2)\ln(s)}{2s}. \tag{49}$$

Recall that the initial values of $s$ and $s^2$ are,

$$s = i\sqrt{\lambda + \omega^2},$$
$$s^2 = -(\lambda + \omega^2). \tag{50}$$

Substituting these into Equation (49) gives,

$$\left.\frac{d\beta}{d\nu}\right|_{\nu=0} = \frac{-\lambda\pi}{4\sqrt{\lambda + \omega^2}} < 0, \quad \text{one possible initial slope,}$$

$$\left.\frac{d\sigma}{d\nu}\right|_{\nu=0} = \frac{\lambda \ln(\lambda + \omega^2)}{4\sqrt{\lambda + \omega^2}}, \quad \text{three possible initial slopes.} \tag{51}$$

Notice that the initial slope for the decay rate is always negative no matter what values of $\lambda$ or $\omega$ are used, as we would expect. The initial slope for the oscillation frequency can be positive, negative, or zero. The sign of the initial slope is determined by $\ln(\lambda + \omega^2)$

$$\lambda + \omega^2 < 1, \quad \text{negative slope,}$$
$$\lambda + \omega^2 = 1, \quad \text{zero slope,} \tag{52}$$
$$\lambda + \omega^2 > 1, \quad \text{positive slope.}$$

Notice that when the initial slope of the oscillation frequency is positive (i.e., when $\lambda + \omega^2 > 1$) this implies that there is a maximum frequency for the oscillator that is greater than $\sigma(\nu = 0) = \sqrt{\lambda + \omega^2}$ and $\sigma(\nu = 2) = \frac{\omega}{\sqrt{1+\lambda}}$ (recall $\frac{\omega}{\sqrt{1+\lambda}} < \sqrt{\lambda + \omega^2}$). The maximum frequency will occur when Equation (47) is zero. Physically this is intriguing. Intuitively it would be guessed that by increasing the order of the derivative, from zero, on the damping term, that the increase in damping would decrease the frequency of oscillation. Equation (52) indicates that this need not be the case.

The other designation is when $\nu = 1$, two possible midpoints for the frequency, three possible midpoints for the decay (see Equation (39)). At $\nu = 1$ for $\lambda > 2\omega$ the decay rate is double valued so a jump discontinuity is expected (see Figures (7.a), (8.a), (9.a), (12.a) and (15.a)). For $\lambda = 2\omega$ consider the slope at $\nu = 1$,

$$\frac{ds}{d\nu} = \frac{s(s^2 + \omega^2)\ln(s)}{s^2 - \omega^2}. \tag{53}$$

Note that in this case $s = -\omega$ so the denominator is zero which means the slope for both the decay rate and the oscillation frequency is undefined (discontinuity or the curve is vertical, see Figures 4, 5, 6, 11, and 14). For $\lambda < 2\omega$ the derivative, Equation (46), at $\nu = 1$, is continuous so a smooth graph is expected for both the decay rate and oscillation frequency.

Consider Equation (46) when $\nu = 2$ (same end point for all curves)

$$\left.\frac{ds}{dv}\right|_{v=2} = \frac{s(s^2 + \omega^2)\ln(s)}{-2\omega^2}. \tag{54}$$

Recall that the final values of $s$ and $s^2$ are,

$$s = i\frac{\omega}{\sqrt{1+\lambda}},$$

$$s^2 = -\frac{\omega^2}{1+\lambda}. \tag{55}$$

Substituting these into Equation (54) gives,

$$\left.\frac{d\beta}{dv}\right|_{v=2} = \frac{\pi}{4}\left(\frac{\lambda\omega}{(1+\lambda)^{3/2}}\right) > 0, \quad \text{one possible final slope,}$$

$$\left.\frac{d\sigma}{dv}\right|_{v=2} = \frac{\lambda\omega}{4(1+\lambda)^{3/2}} \ln\left(\frac{1+\lambda}{\omega^2}\right), \quad \text{three possible final slopes.} \tag{56}$$

Notice that the final slope for the decay rate is always positive no matter what values for $\lambda$ or $\omega$ are used, as would be expected. The effective decay rate goes to zero as the damping disappears. The final slope for the oscillation frequency can be positive, negative or zero. The sign of the final slope is determined by $ln\left(\frac{1+\lambda}{\omega^2}\right)$;

$$\begin{aligned} 1 + \lambda &> \omega^2 \quad \text{positive final slope,} \\ 1 + \lambda &= \omega^2 \quad \text{zero final slope,} \\ 1 + \lambda &< \omega^2 \quad \text{negative final slope.} \end{aligned} \tag{57}$$

Intuitively a positive, or even zero, slope for the effective oscillation frequency might be expected. The negative slope is not. Of the three initial slope options for the frequency, three types of roots for the resulting quadratic equation at $v = 1$, and the three final slope options for the frequency 15 types of solution curves are possible.

**Table 1.** List of possible types of solutions for Equation (34)

| Initial Slope | Quadratic Type | Final Slope | Curves | Sample $\lambda$ value | Sample $\omega^2$ value |
|---|---|---|---|---|---|
| + | $\lambda < 2\omega$ | + | Figures 1 | 1 | 9/16 |
| + | $\lambda < 2\omega$ | 0 | Figures 2 | 1/2 | 3/2 |
| + | $\lambda < 2\omega$ | - | Figures 3 | 1 | 4 |
| + | $\lambda = 2\omega$ | + | Figures 4 | 1 | 1/4 |
| + | $\lambda = 2\omega$ | 0 | Figures 5 | $2 + 2\sqrt{2}$ | $(1 + \sqrt{2})^2$ |
| + | $\lambda = 2\omega$ | - | Figures 6 | 6 | 9 |
| + | $\lambda > 2\omega$ | + | Figures 7 | 3 | 1 |
| + | $\lambda > 2\omega$ | 0 | Figures 8 | 15 | 16 |
| + | $\lambda > 2\omega$ | - | Figures 9 | 9 | 16 |
| 0 | $\lambda < 2\omega$ | + | Figures 10 | 1/2 | 1/2 |
| 0 | $\lambda < 2\omega$ | 0 | No | | |
| 0 | $\lambda < 2\omega$ | - | No | | |
| 0 | $\lambda = 2\omega$ | + | Figures 11 | $2\sqrt{2} - 2$ | $(\sqrt{2} - 1)^2$ |
| 0 | $\lambda = 2\omega$ | 0 | No | | |
| 0 | $\lambda = 2\omega$ | - | No | | |
| 0 | $\lambda > 2\omega$ | + | Figures 12 | 15/16 | 1/16 |
| 0 | $\lambda > 2\omega$ | 0 | No | | |
| 0 | $\lambda > 2\omega$ | - | No | | |
| - | $\lambda < 2\omega$ | + | Figures 13 | 1/4 | 1/4 |
| - | $\lambda < 2\omega$ | 0 | No | | |
| - | $\lambda < 2\omega$ | - | No | | |
| - | $\lambda = 2\omega$ | + | Figures 14 | 2/3 | 1/9 |
| - | $\lambda = 2\omega$ | 0 | No | | |
| - | $\lambda = 2\omega$ | - | No | | |
| - | $\lambda > 2\omega$ | + | Figures 15 | 7/8 | 1/16 |
| - | $\lambda > 2\omega$ | 0 | No | | |
| - | $\lambda > 2\omega$ | - | No | | |

The graphs of the possible curves for the listed sample values are listed after the References below. The first nine pairs of graphs (decay rate and the oscillation frequency) show the afore mentioned interesting feature for fractional oscillators, the initial slope of the oscillation frequency is positive, hence, there is a maximum frequency for the oscillator that is greater than $\sigma(\nu = 0) = \sqrt{\lambda + \omega^2}$ and $\sigma(\nu = 2) = \frac{\omega}{\sqrt{1+\lambda}}$. The mathematical reason for this is clear; the physical reason for this is not.

As can be seen in graphs (7.a), (8.a), (9.a), (12.a), and (15.a) there are jump discontinuities in the decay rate. These are the cases for which $\nu = 1$ and $\lambda > 2\omega$ causing there to be two real roots for the decay rate

$$\beta = \frac{-\lambda \pm \sqrt{\lambda^2 - 4\omega^2}}{2}. \tag{58}$$

Looking at the graphs for these cases it can be seen that

$$\lim_{\nu \to 1^-} \beta = \frac{-\lambda - \sqrt{\lambda^2 - 4\omega^2}}{2}, \tag{59}$$

$$\lim_{\nu \to 1^+} \beta = \frac{-\lambda + \sqrt{\lambda^2 - 4\omega^2}}{2}. \tag{60}$$

$\nu = 1$ is analogous to a critical temperature for a substance with two phases (c.f. page 35 of Sears and Salinger (1975)).

Taking the point of view that Equation (34) is an Equation of state with $\beta$ and $\sigma$ being the dependent variables and $\nu$, $\omega$, and $\lambda$ being the independent variables ($\nu$ would be analogous to temperature or perhaps an applied external field). Figures 1, 2, 3, 10, and 13 would represent systems smoothly transitioning as $\nu$ ranges from 0 to 2. The decay rate starts and ends at zero with a single minimum occurring for some $\nu$ value that is not at the endpoints. The behavior for the frequency is more complicated but still smooth. Figures 1.b, 2.b, and 3.b all have a maximum that is not on an endpoint. Figures 1.b, 10.b, and 13.b all have a minimum that is not on an endpoint. Figures 4, 5, 6, 11, and 14 have two additional features: the decay rate curve has an inflection point at the point where the slope becomes vertical ($\nu = 1$); this is also the same point where the frequency has a kink. At the kink the frequency rapidly goes to zero, from either side, giving the appearance of a lambda transition, albeit upside down, (c.f. pg. 193 of Sears and Salinger (1975)). Figures 7, 8, 9, and 12 not only have a kink in the frequency graphs but also a jump discontinuity in the decay rate graphs (at $\nu = 1$). For these curves the decay rate falls smoothly for $0 \leq \nu < 1$. At $\nu = 1$ the decay rate has two different values, two states existing in equilibrium as it were. For $1 < \nu \leq 2$ the decay rate goes back to being a smooth function, though not necessarily monotonic (e.g. Figures 7.a and 12.a).

## 6. Conclusion

In this paper a generalized Bagley-Torvik Equation was solved using the Laplace transform. The solution was found to be expressible in terms sine, cosine, and the exponential function and their convolutions. The advantage of expressing the solution in this fashion, instead of a series of Mittag-Leffler functions, is that the behavior of the solution is readily apparent (decaying oscillations of some sort).

The effective decay rate and oscillation frequency were also examined. These arose out of a solution to a transcendental Equation. The dependence of the effective decay rate and oscillation frequency on $\nu, \lambda$, and $\omega$ was found to be analogous to a thermodynamic Equation of state. Varying $\nu$ was found to cause behavior akin to a phase change in some cases and lambda transitions in others. In that sense changing the variable $\nu$ could be viewed as having a similar effect to changing the temperature (or an external field) in a thermodynamic system.

It is hoped that other researchers will be able to set up a "Generalized Bagley-Torvik oscillator" with physical fluids whose $v$ value is different than $3/2$ so that the physical reasons for the variety of graphs for the effective decay rate and oscillation frequency can be understood.

## REFERENCES

Bagley, R. L. and Torvik, P. J. (1984). On the appearance of the fractional derivative in the behavior of real materials, J. Appl. Mech., 51, pp. 294-298.

Bansal, M. K. and Jain, R. (2016) Analytical Solution of Bagley Torvik Equation by Generalize Differential Transform, Int. Jour. of Pure and Applied Mathematics, Vol. 110, No. 2, 265-273.

Burov, S. and Barkai, E. (2007). The Critical Exponent of the Fractional Langevin Equation is $\alpha_c \approx 0.402$, arXiv:0712.3407v1 [cond-mat.stat-mech] 20 Dec 2007.

Burov, S. and Barkai, E. (2008). Fractional Langevin Equation: Over-Damped, Under-Damped, and Critical behaviors, arXiv:0802.3777v1 [cond-mat.stat-mech] 26 Feb.

Diethelm, K. and Ford, J. (2002). Numerical Solution of the Bagley-Torvik Equation, BIT Numerical Mathematics, Vol.42, Num. 3, Sept.

Galucio, A. C., Dubois, J. F., and Dubois, F. (2006). On the use of fractional derivative operators to describe viscoelastic damping in structural dynamics- FE formulation of sandwich beams and approximation of fractional derivatives by using the $G^\alpha$ scheme, Derivation fractionaire en mecanique – Etat-de-l'art et applications, CNAM Paris – 17$^{th}$ November, 2006. http://www.cnam.fr/lmssc/seminaires/derivfrac/galucio/17NOV.PDF

Koh, C. G. and Kelly, J. M. (1990). Application of Fractional Derivatives to Seismic Analysis of Base-Isolated Models, Earthquake Engineering and Structural Dynamics, Vol 19, 229-241.

Labecca, W., Guimarães, O., and Piqueira, J. R. C. (2015). Analytical Solution of General Bagley-Torvik Equation, Mathematical Problems in Engineering, vol. 2015, Article ID 591715, 4 pages. https://doi.org/10.1155/2015/591715.

Naber, M. (2010). Linear Fractional Damped Oscillator, Int. Jour. of Dif. Equation Vol. 2010, Special Issue on Fractional Differential Equations, Article ID 186928.

Narahari Achar, B. N., Hanneken, J. W., and Clarke, T. (2004). Damping characteristics of a fractional oscillator, Physica A: Statistical Mechanics and its Applications,Vol. 339, issues 3-4, 15 Aug 2004, pages 311-319.

Podlubny, I. (1999). *Fractional Differential Equations*, Academic Press.

Rajagopal, K. R. and Na, T. Y. (1983). On the Stokes' Problem for a Non-Newtonian Fluid, Acta Mechanica 48, 233-239.

Ray, S. S. and Bera, R. K. (2005). Analytical solutions of the Bagley Torvik Equation by Adomian decomposition method, Appl. Math. and Comp., vol 168, issue 1, Sept. 2005, pp. 398-410.

Sears, Francis W. and Salinger, Gerhard L. (1975). *Thermodynamics, Kinetic Theory, and Statistical Thermodynamics*, 3$^{rd}$ edition, Addison-Wesley.

Zaslavsky, G. M., Stanislavsky, A.A. and Edelman, M. (2005). Chaotic and Pseudochaotic Attractors of Perturbed Fractional Oscillator, arXiv:nlin.CD/0508018 v1 10 Aug 2005.

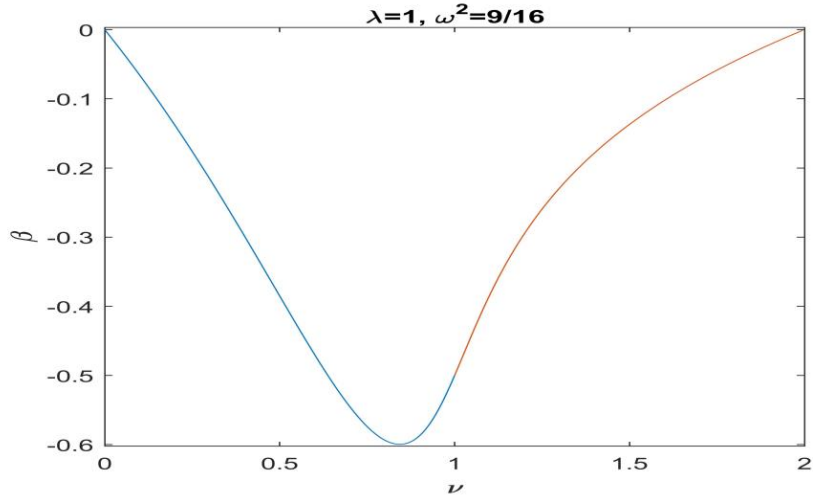

**Figure 1.a.**  This is a graph of the decay rate with derivative order.

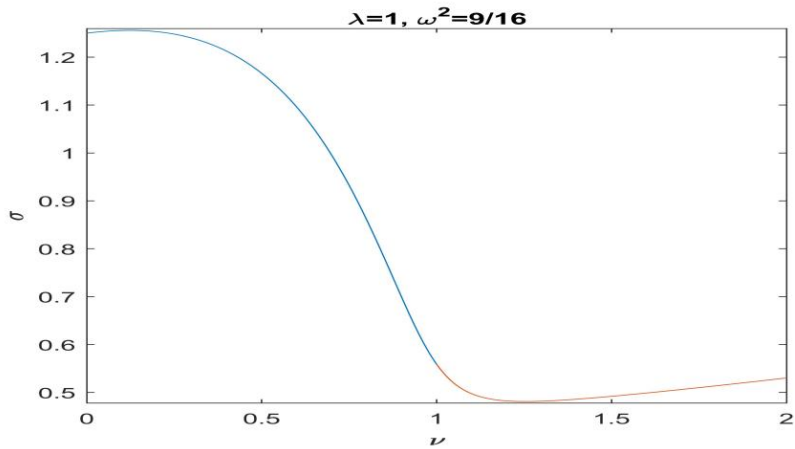

**Figure 1.b.**  This is a graph of the frequency with derivative order.

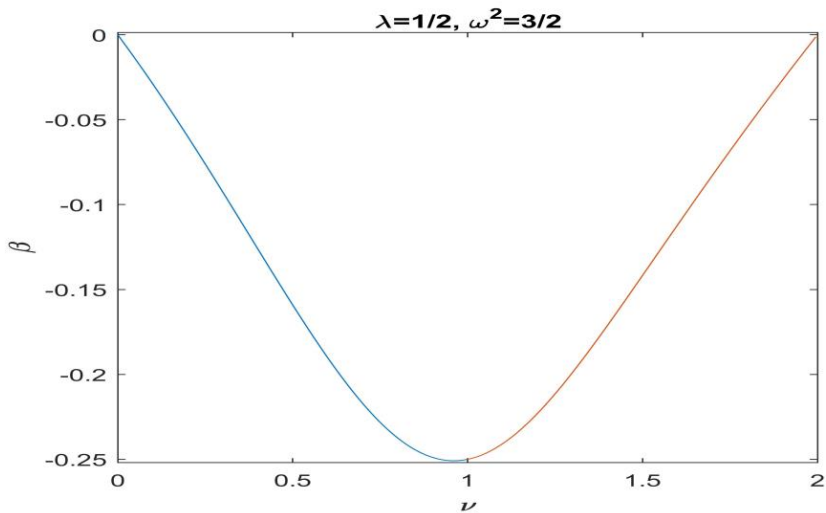

**Figure 2.a.**  This is a graph of the decay rate with derivative order.

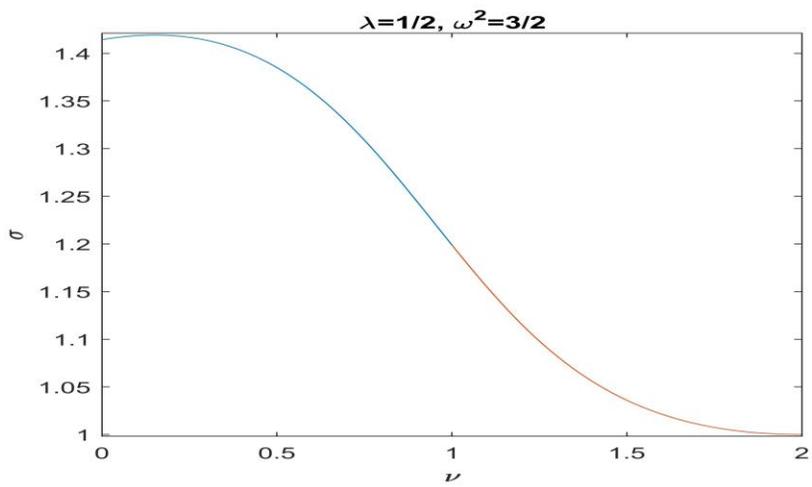

**Figure 2.b.** This is a graph of the frequency with derivative order.

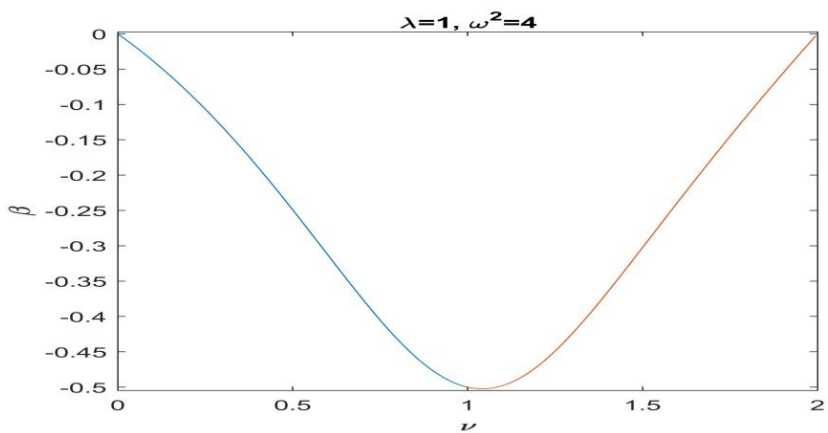

**Figure 3.a.** This is a graph of the decay rate with derivative order.

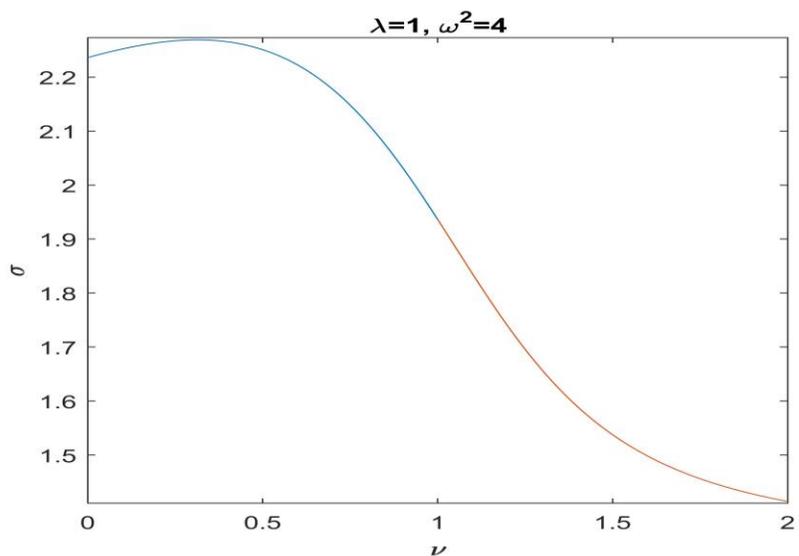

**Figure 3.b.** This is a graph of the frequency with derivative order.

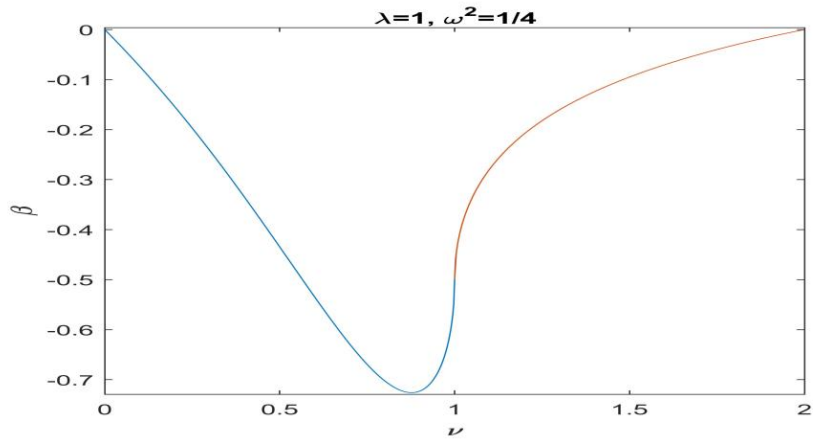

**Figure 4.a.** This is a graph of the decay rate with derivative order.

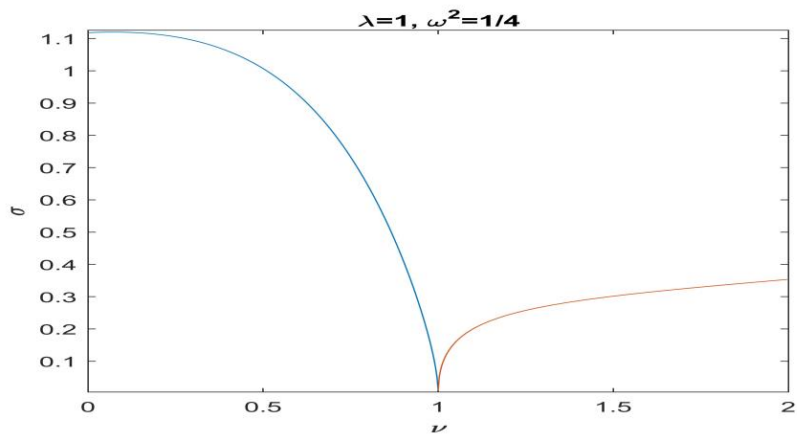

**Figure 4.b.** This is a graph of the frequency with derivative order.

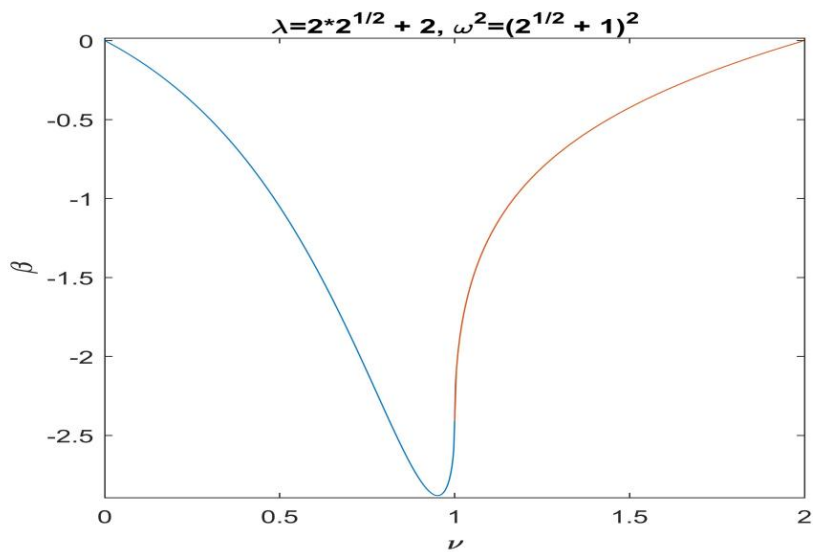

**Figure 5.a.** This is a graph of the decay rate with derivative order.

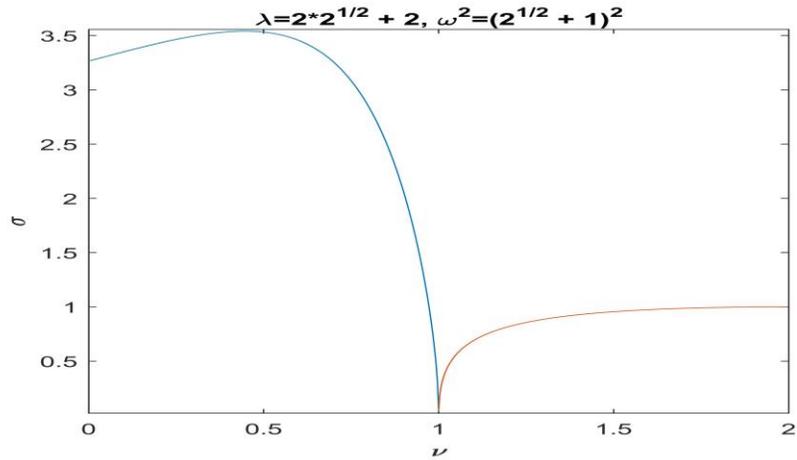

**Figure 5.b.** This is a graph of the frequency with derivative order.

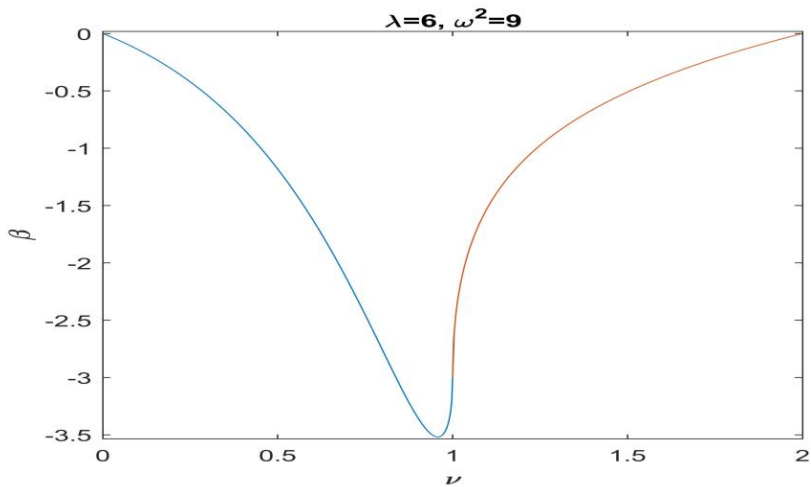

**Figure 6.a.** This is a graph of the decay rate with derivative order.

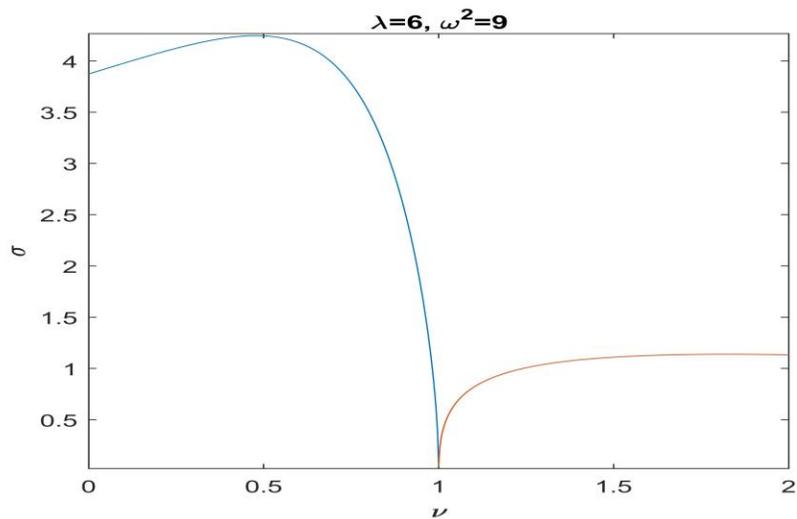

**Figure 6.b.** This is a graph of the frequency with derivative order.

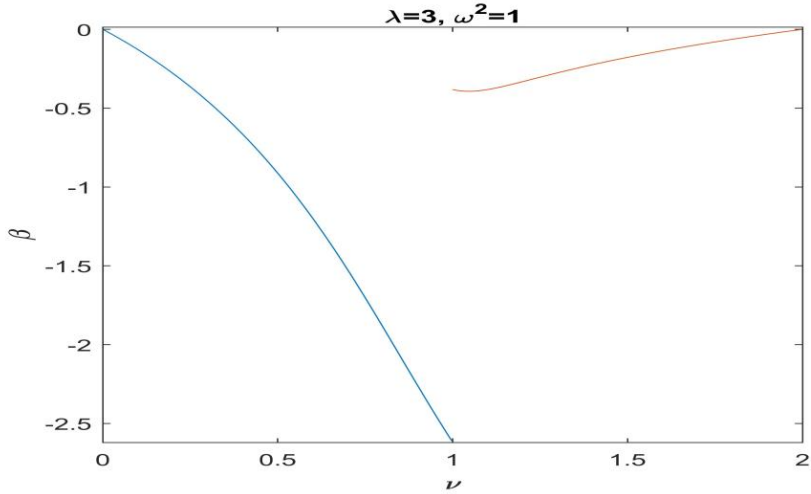

**Figure 7.a.** This is a graph of the decay rate with derivative order.

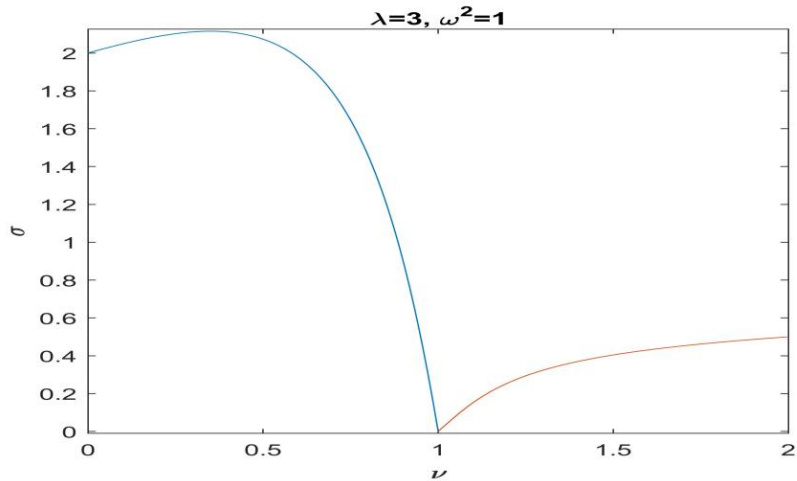

**Figure 7.b.** This is a graph of the frequency with derivative order.

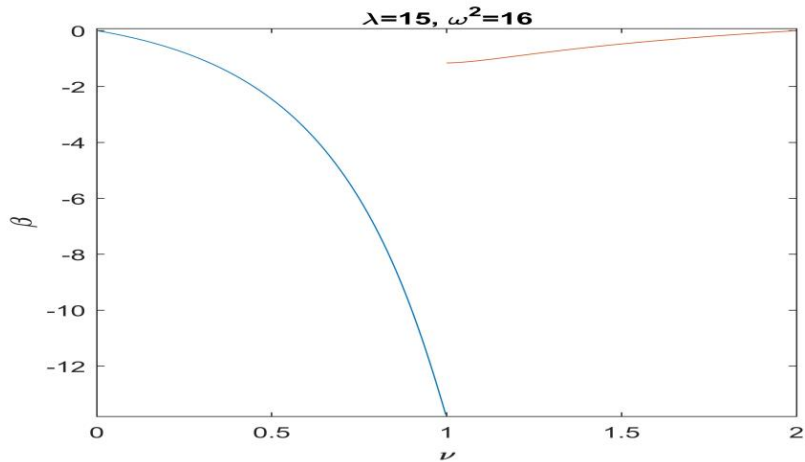

**Figure 8.a.** This is a graph of the decay rate with derivative order.

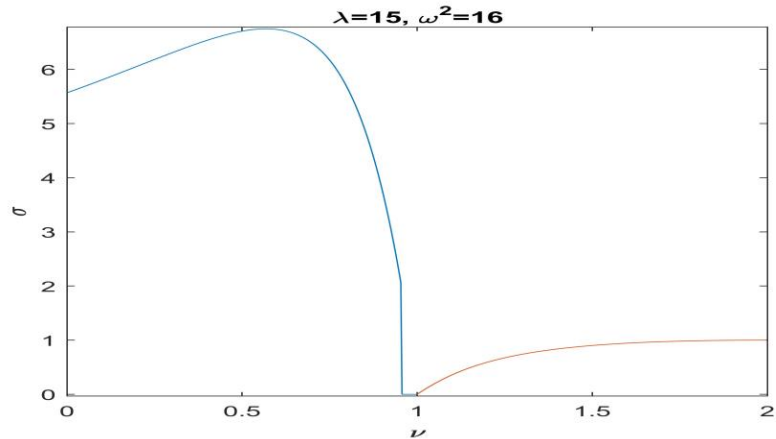

**Figure 8.b.** This is a graph of the frequency with derivative order.

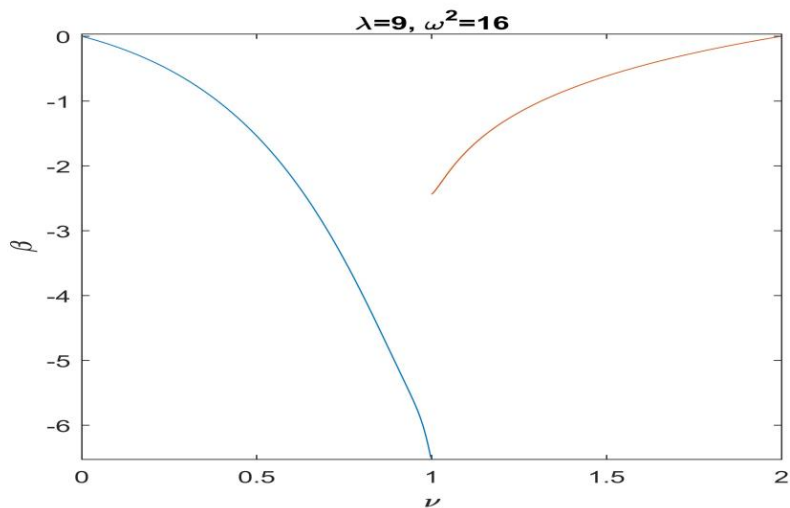

**Figure 9.a.** This is a graph of the decay rate with derivative order.

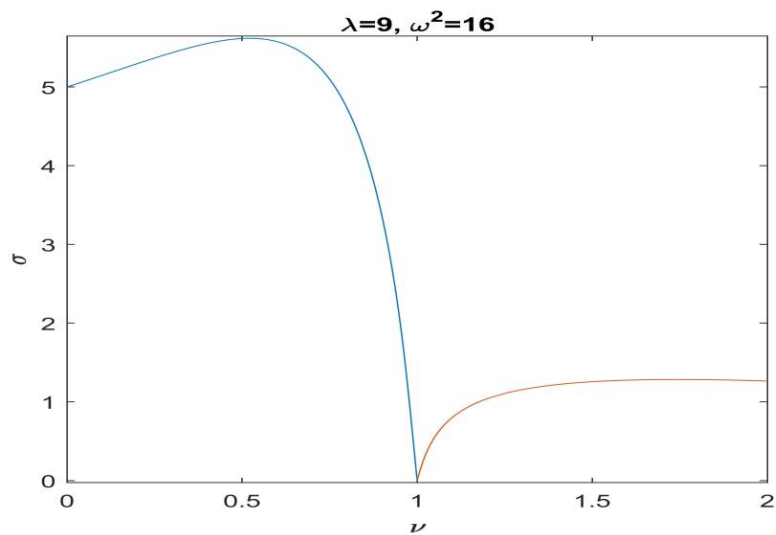

**Figure 9.b.** This is a graph of the frequency with derivative order.

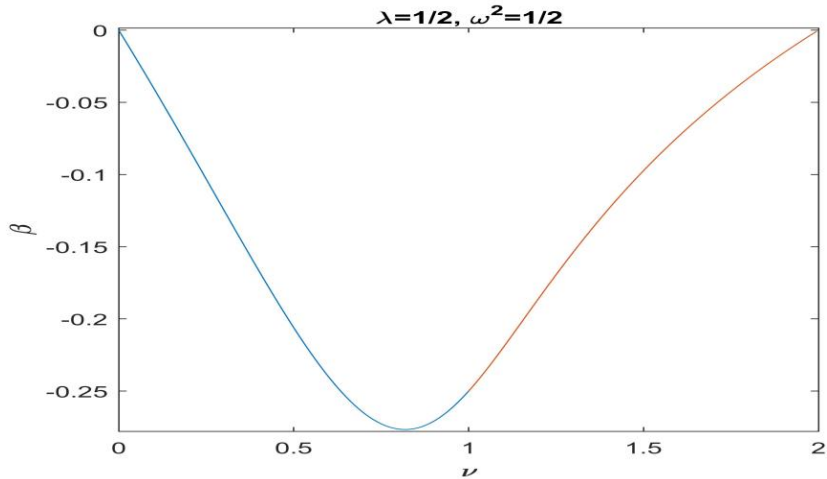

**Figure 10.a.** This is a graph of the decay rate with derivative order.

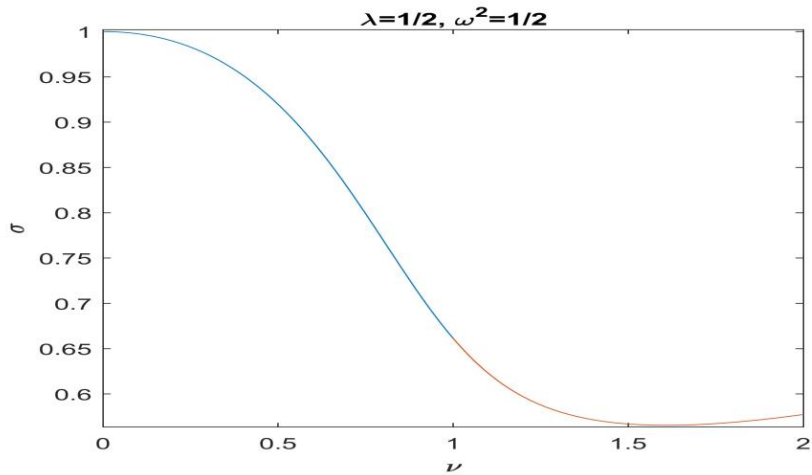

**Figure 10.b.** This is a graph of the frequency with derivative order.

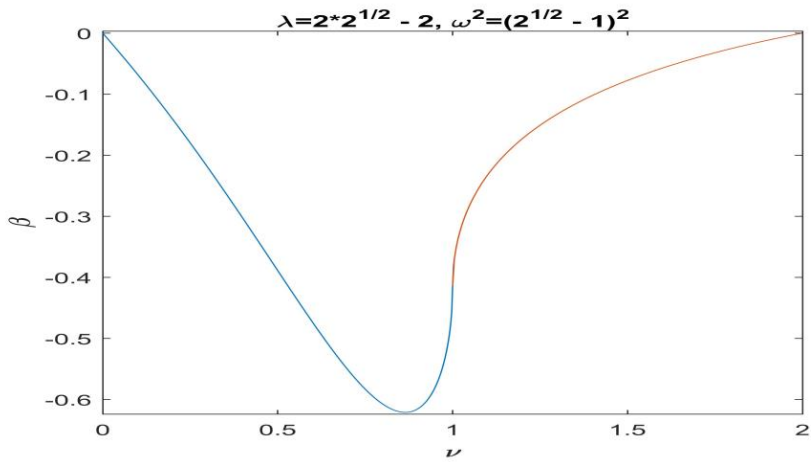

**Figure 11.a.** This is a graph of the decay rate with derivative order.

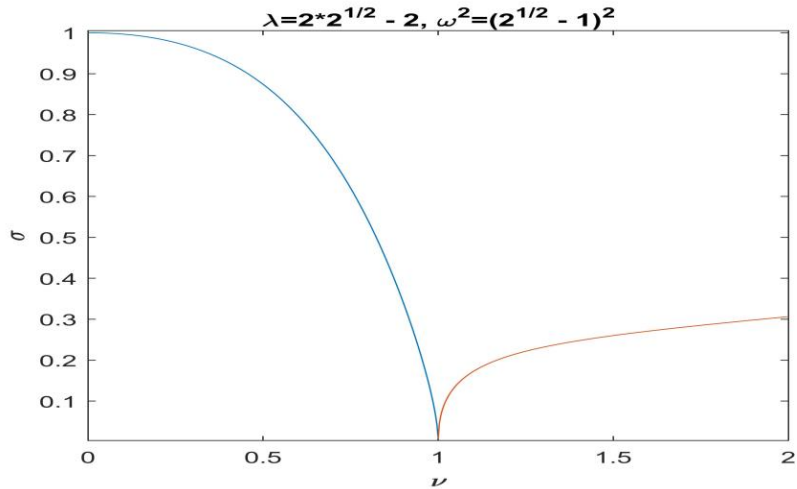

**Figure 11.b.** This is a graph of the frequency with derivative order.

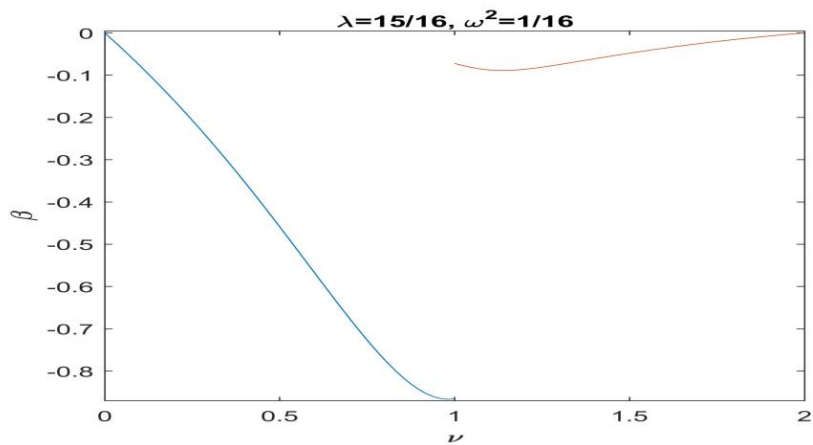

**Figure 12.a.** This is a graph of the decay rate with derivative order.

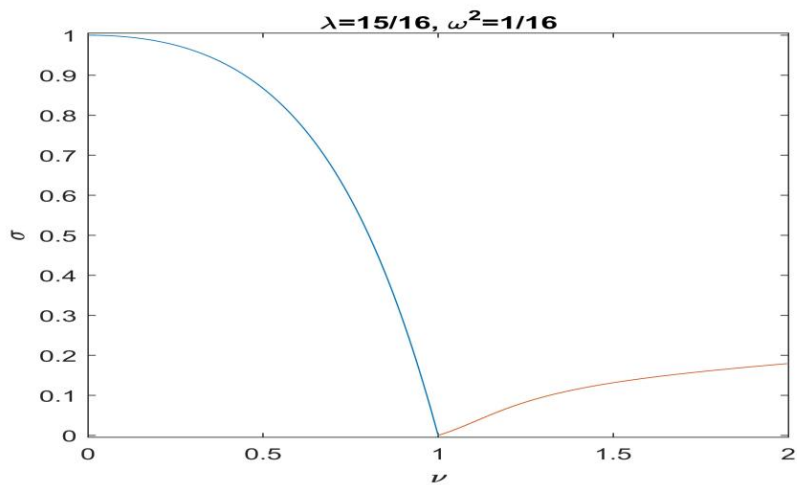

**Figure 12.b.** This is a graph of the frequency with derivative order.

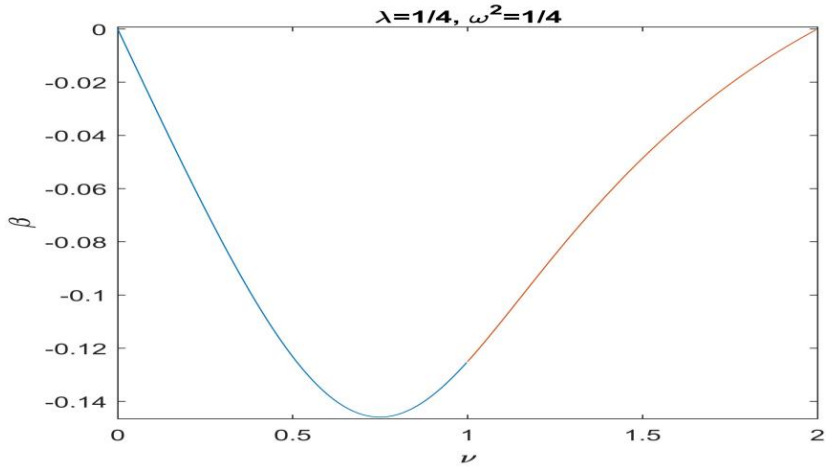

**Figure 13.a.** This is a graph of the decay rate with derivative order.

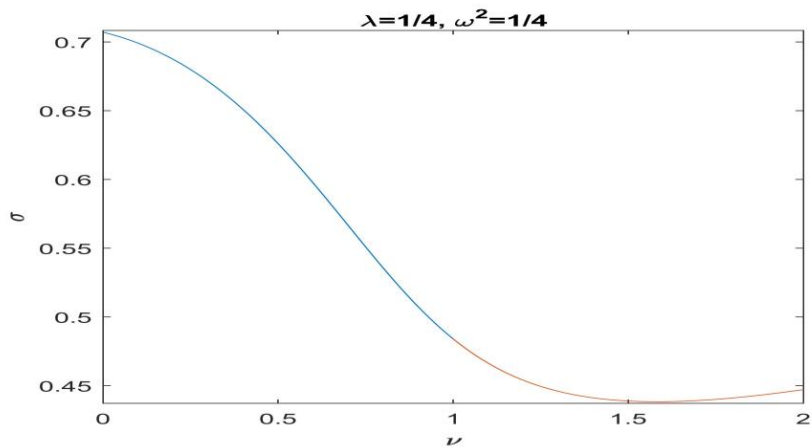

**Figure 13.b.** This is a graph of the frequency with derivative order.

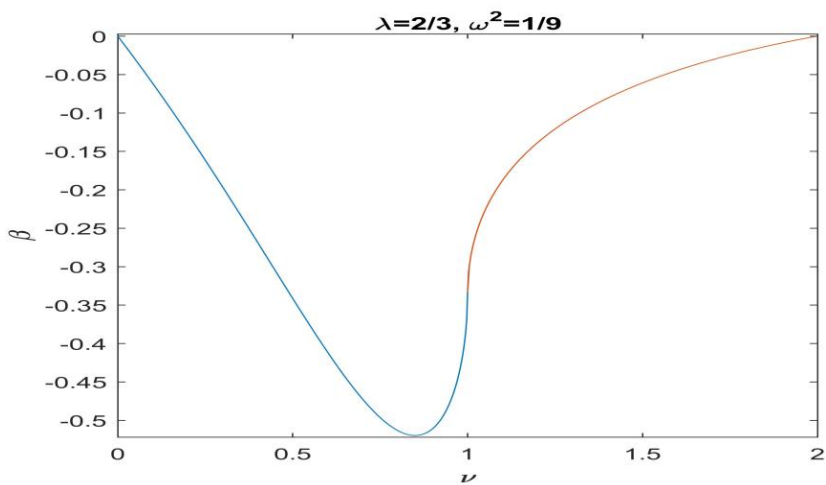

**Figure 14.a.** This is a graph of the decay rate with derivative order.

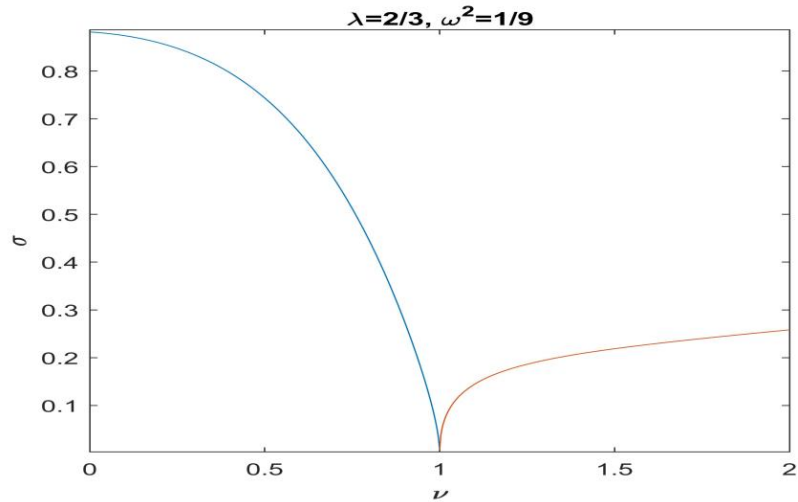

**Figure 14.b.** This is a graph of the frequency with derivative order.

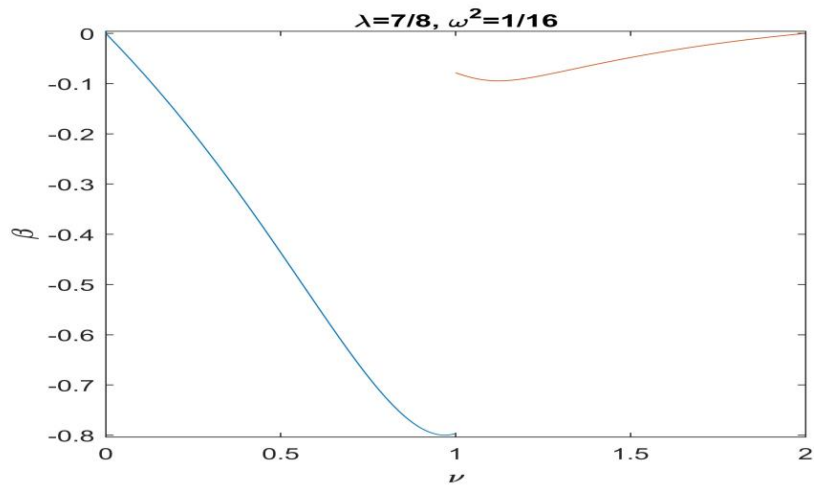

**Figure 15.a.** This is a graph of the decay rate with derivative order.

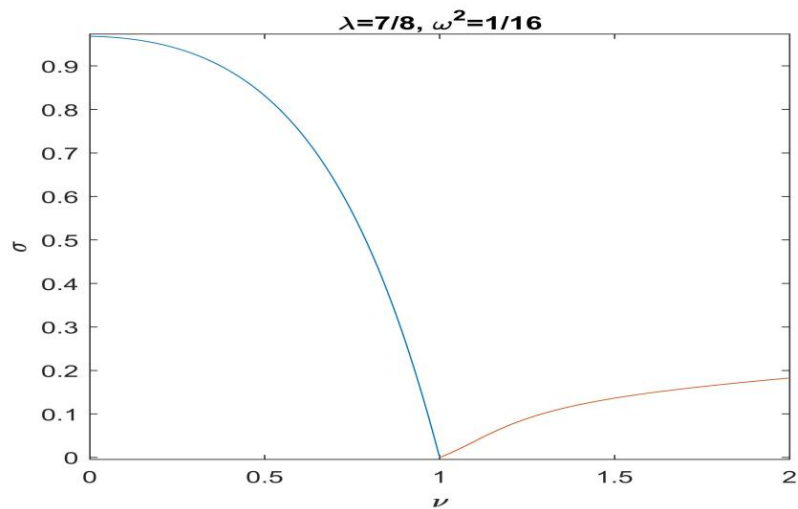

**Figure 15.b.** This is a graph of the frequency with derivative order.